\documentclass[%
reprint,
superscriptaddress,
amsmath,amssymb,
aps,
pra,
longbibliography]{revtex4-1}

\usepackage{graphicx}
\usepackage{mathtools} 
\usepackage{textcomp} 
\usepackage[space]{grffile} 
\usepackage{dcolumn}
\usepackage{bm}
\usepackage{textgreek} 

\newcommand{\ket}[1]{\left| #1 \right>} 
\newcommand{\ketbra}[2]{{\left|#1\right>\hspace{-3pt}\left<#2\right|}} 

\begin{document}


\title{Individual Control and Readout of Qubits in a Sub-Diffraction Volume}

\author{Eric Bersin}
\thanks{These two authors contributed equally.}
\affiliation{Department of Electrical Engineering and Computer Science, Massachusetts Institute of Technology, Cambridge, Massachusetts 02139, USA}

\author{Michael Walsh}
\thanks{These two authors contributed equally.}
\affiliation{Department of Electrical Engineering and Computer Science, Massachusetts Institute of Technology, Cambridge, Massachusetts 02139, USA}

\author{Sara L. Mouradian}
\affiliation{Department of Electrical Engineering and Computer Science, Massachusetts Institute of Technology, Cambridge, Massachusetts 02139, USA}

\author{Matthew E. Trusheim}
\affiliation{Department of Electrical Engineering and Computer Science, Massachusetts Institute of Technology, Cambridge, Massachusetts 02139, USA}

\author{Tim Schr\"{o}der}
\thanks{Current address: Niels Bohr Institute, University of Copenhagen, Blegdamsvej 17, 2100 Copenhagen, Denmark}
\affiliation{Department of Electrical Engineering and Computer Science, Massachusetts Institute of Technology, Cambridge, Massachusetts 02139, USA}

\author{Dirk Englund}
\email{englund@mit.edu}
\affiliation{Department of Electrical Engineering and Computer Science, Massachusetts Institute of Technology, Cambridge, Massachusetts 02139, USA}

\begin{abstract}
Medium-scale ensembles of coupled qubits offer a platform for near-term quantum technologies including computing~\cite{Preskill_2018}, sensing~\cite{Degen_2017}, and the study of mesoscopic quantum systems~\cite{Zhang_2017,Choi_2017,Mohammady_2018}. Atom-like emitters in solids~\cite{Gao_2015} have emerged as promising quantum memories, with demonstrations of spin-spin entanglement by optical~\cite{Humphreys_2018} and magnetic~\cite{Dolde_2013} interactions. Magnetic coupling in particular is attractive for efficient and deterministic entanglement gates, but raises the problem of individual spin addressing at the necessary nanometer-scale separation. Current super-resolution techniques~\cite{Chen_2013,Jaskula_2017} can reach this resolution, but are destructive to the states of nearby qubits. Here, we demonstrate the measurement of individual qubit states in a sub-diffraction cluster by selectively exciting spectrally distinguishable nitrogen vacancy (NV) centers~\cite{Doherty_2013}. We demonstrate super-resolution localization of single centers with nanometer spatial resolution, as well as individual control and readout of spin populations. These measurements indicate a readout-induced crosstalk on non-addressed qubits below $4\times10^{-2}$. This approach opens the door to high-speed control and measurement of qubit registers in mesoscopic spin clusters, with applications ranging from entanglement-enhanced sensors~\cite{Tanaka_2015} to error-corrected qubit registers~\cite{Waldherr_2014,Cramer_2017} to multiplexed quantum repeater nodes~\cite{Sinclair_2014,van_Dam_2017}.
\end{abstract}
\maketitle

\begin{figure*}
\centering
\includegraphics{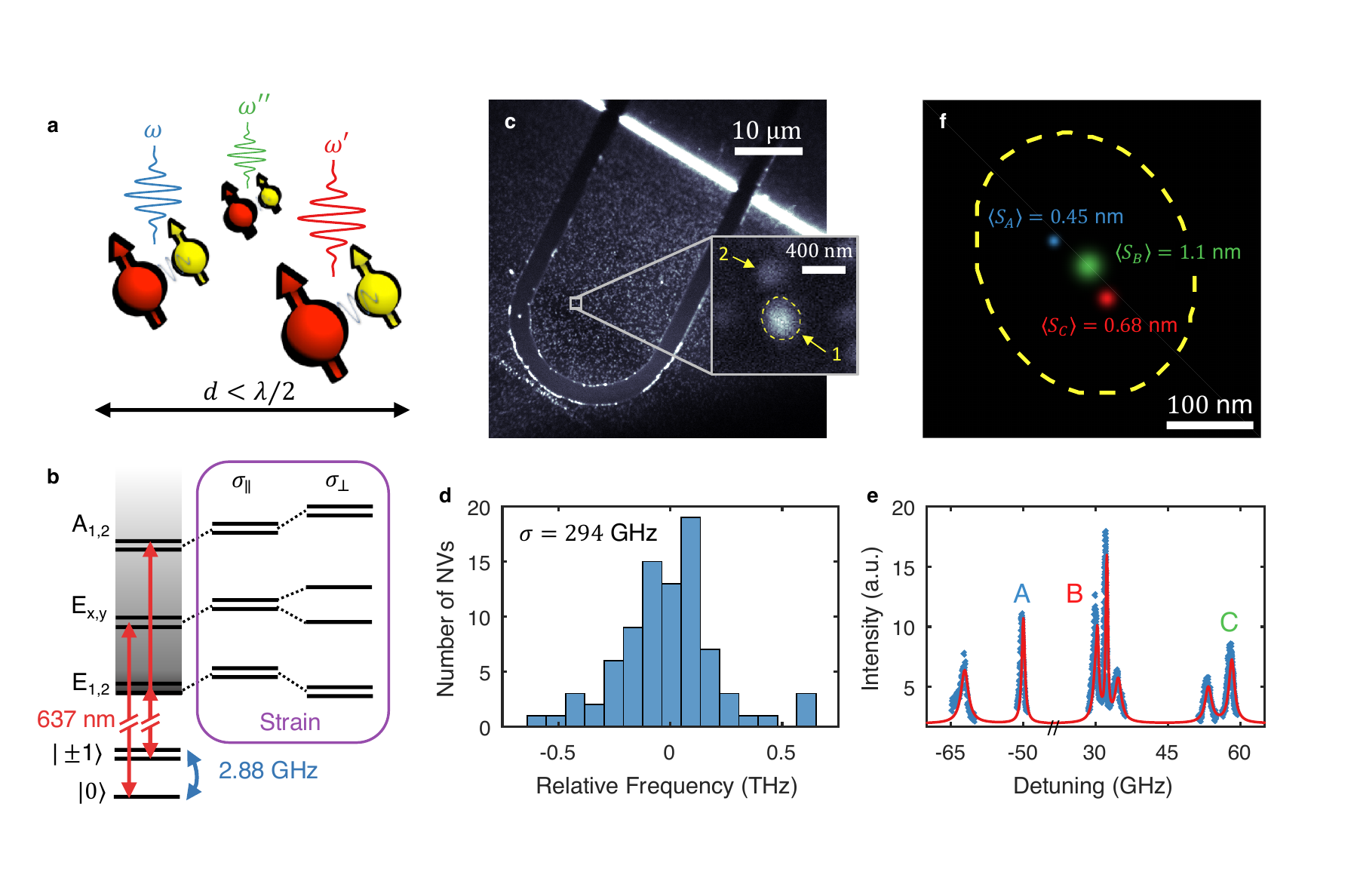}
\vspace{-20pt}
\caption{\textbf{Preferential excitation and imaging of sub-diffraction defects.} \textbf{a}, An ensemble of NV electron spins (red) coupled to nearby nuclear spins (yellow) with distinct optical transition frequencies due to local strain fields, allowing selective interaction with individual centers operating on different frequency channels within a diffraction-limited spot. \textbf{b}, NV electron level structure. A ground state spin triplet acts as a qubit, addressable via 2.88~GHz microwave driving (blue). The spin-conserving radiative transitions (red) are at distinct frequencies, allowing optical spin readout via resonant excitation. Axial ($\sigma_{\parallel}$) and transverse ($\sigma_{\perp}$) strains shift these levels further and may distinguish individual NVs. \textbf{c}, Confocal microscope image of the PCD showing the microwave stripline. The bright white line is a grain boundary in the diamond. Inset, a close-up scan showing a brighter spot at site 1, determined to be a cluster of NVs, as well as a dimmer spot at site 2, determined to be a single NV. \textbf{d}, Histogram (N=87) showing the inhomogeneous distribution of ZPL transitions measured via PLE in the PCD, with a standard deviation $\sigma=294$~GHz. \textbf{e}, PLE on the NV cluster at site 1, showing multiple transitions from strain-split centers, fit to a sum of 7 Lorentzians (red). \textbf{f}, Reconstructed locations within the cluster, with width (standard deviation) indicating the standard distance error on each point, multiplied 10x for visibility. For comparison, the dashed line shows the full-width half-maximum size of the spot under 532~nm excitation.}
\label{fig:1}
\end{figure*}

Major advances towards larger coherent spin systems in diamond have recently been made by controlling nuclear spins~\cite{Abobeih_2018} or dark electron spins~\cite{Rosenfeld_2017} through one NV center and by coupling electron spins of two NV centers~\cite{Dolde_2013}. These advances suggest that a system of strongly coupled color centers, each coupled to proximal dark spins, could provide a scalable platform for controlled spin-spin interactions, as illustrated in Figure~\ref{fig:1}a. 

A central challenge for coherent control of such multi-spin systems is the ability to measure individual NV centers without collapsing the states of nearby NVs. Here we address this problem by making use of the inhomogeneous distribution of NV center optical transitions, attributed to natural and defect-induced lattice strain~\cite{Bernien_2012,Sipahigil_2014}. The strain field $\vec{\sigma}$ enters the excited state Hamiltonian as ${H_{\text{strain}}=\vec{\sigma}\cdot\vec{V}}$~\cite{Doherty_2011}, where $\vec{V}$ is the vector of orbital operators, and can be divided into axial and transverse components, with differing effects detailed in Figure~\ref{fig:1}b. We find that this distribution persists even for closely-spaced NV centers, allowing us to optically address individual emitters within a diffraction-limited volume.

We investigate this approach to multi-qubit readout in a Type IIa polycrystalline diamond (PCD, see Methods). The scanning confocal image in Figure 1c shows NV centers in one domain of this PCD near a gold stripline for microwave delivery that cuts across a grain boundary, visible as the bright strip in the image. 
Despite the high strain of the PCD~\cite{Trusheim_2016}, its low nitrogen content allows for NVs with coherence times exceeding 200~\textmu s~\cite{supplement} at room temperature. The histogram of the NV optical transitions (Figure 1d) indicates an inhomogeneous distribution with standard deviation of 294 GHz, nearly 5 times broader than what we measure in single-crystal diamond samples~\cite{supplement}.

This broad inhomogeneous distribution allows us to spectrally distinguish NVs below the diffraction limit. Figure~\ref{fig:1}e shows a photoluminescence excitation (PLE) spectrum taken on a representative fluorescence site on the sample, labeled site 1 in the inset of Figure~\ref{fig:1}c. The spectrum reveals several distinct zero-phonon line (ZPL) peaks, indicative of the presence of multiple NV centers within the diffraction-limited spot. Spatially scanning a narrow laser resonant with one of the transitions in Figure~\ref{fig:1}e preferentially induces fluorescence from a single NV center, selectively imaging this defect out of the cluster. Performing this scan for each observed transition, we find that they correspond to only three spatial positions. Second-order autocorrelation and optically detected magnetic resonance (ODMR) measurements further confirm the presence of three NV centers in this site~\cite{supplement}. The most prominent and well-isolated peaks for each NV center are labeled as A, B, and C in Figure~\ref{fig:1}e. For these transitions, we repeat the resonant imaging experiment, each time fitting the result with a Gaussian point-spread function. The standard error on the fit centers gives a localization precision of ${\langle S_A\rangle = 0.45}$~nm for the brightest and most spectrally distinct NV~\cite{supplement}. Figure~\ref{fig:1}f shows the reconstructed positions, with spot widths indicating 10 times the localization precision after 40 minutes of integration and the dashed overlay showing the full-width half-maximum size of the original diffraction-limited spot.

\begin{figure*}
\centering
\includegraphics{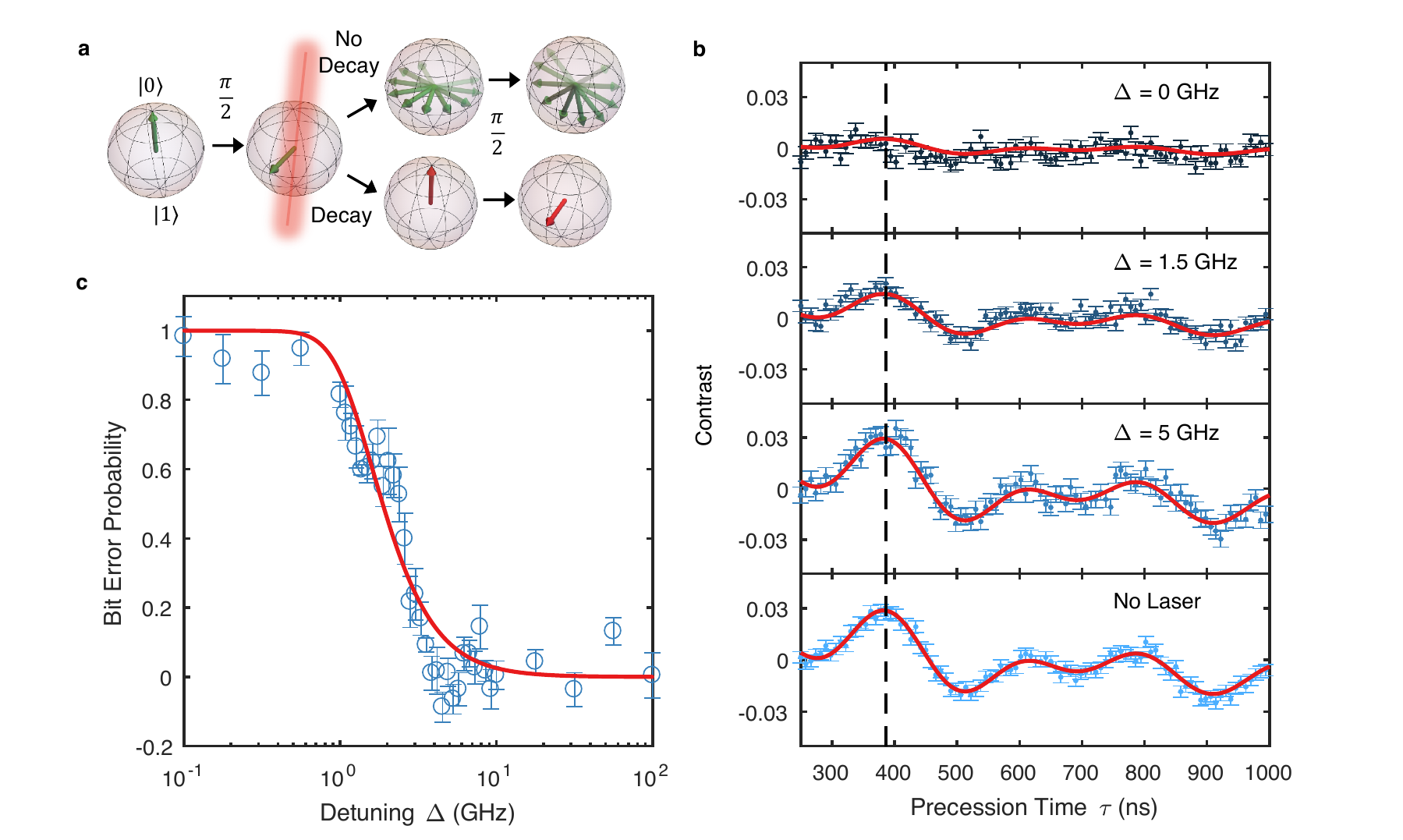}
\caption{\textbf{Qubit degradation under near-resonant excitation for a single isolated NV.} \textbf{a}, Bloch sphere schematic for crosstalk measurement sequence. After a $\pi/2$-pulse prepares the qubit in a superposition state, it precesses around the Bloch sphere during application of a near-resonant laser. With probability $(1-\Gamma)$, the laser will not induce excitation and subsequent decay, preserving the phase built up during the precession period, which is then mapped into population by a second $\pi/2$-pulse and read out with a non-resonant readout pulse. However, the laser may also induce a spin-projecting decay event; in this scenario, the second $\pi/2$-pulse will always place the spin in an even superposition state, independent of any phase accumulated during the precessionary period, leading to a precession time-independent intensity at the final readout. \textbf{b}, Ramsey sequences with a resonant laser of varying detunings applied during the free precession period. The fits (solid lines) have one fit parameter for the fringe amplitude relative to that of a reference Ramsey taken with no crosstalk laser. \textbf{c}, Crosstalk probability as a function of laser detuning, taken by fixing the precession time to the fringe maximum at $\tau=386$~ns (dashed line in \textbf{b}) and sweeping the resonant laser detuning. The contrast values are normalized to the fringe amplitude from the no-laser case. In red, the model for $\Gamma$ (Eq. \ref{eq:crosstalk}) with one fit parameter for the optical Rabi frequency.}
\label{fig:2}
\end{figure*}

We next consider the crosstalk that an optical readout of one NV induces in other NVs in a diffraction-limited spot. For simplicity, we first study these dynamics in a simple spin-1 system associated with a single NV center (NV$_\text{D}$) in site 2 of Figure 1c, which is initialized into state $\ket{\psi_0}=\ket{m_s=0}+\ket{m_s=1}$. Suppose a laser is applied at frequency $\omega_L$ for time $T$ to perform resonant readout on a hypothetical neighboring NV. This laser non-resonantly excites NV$_\text{D}$ from ground state $\ket{i}$ into excited state $\ket{k}$, projecting its state by spontaneous emission into ground state $\ket{j}$, where $i,j\in m_s=\{-1,0,1\}$ and $k\in\{E_1,E_2,E_x,E_y,A_1,A_2\}$, with probability~\cite{supplement}:
\begin{equation}
\Gamma_{ijk}=1-\exp\left(-\frac{\gamma_{jk} \Omega_{ij}^{2} T}{2(\Omega_{ij}^2 + \Delta_{ij}^2)}\right),
\label{eq:crosstalk}
\end{equation}
where $\Delta_{ij}$ is the detuning of $\omega_L$ from NV$_\text{D}$'s $\ket{i}\rightarrow\ket{j}$ ground-to-excited state transition, $\Omega_{ij}$ is the optical Rabi frequency, and $\gamma_{jk}$ is the excited state's decay rate into $\ket{k}$. In addition to such a spontaneous-emission-induced state projection, NV$_\text{D}$ may also acquire a phase shift due to the AC-Stark shift of the applied laser; however, this is a weak and coherent process and can be compensated~\cite{supplement}.

We probe this laser-induced crosstalk using Ramsey interferometry, as illustrated in Figure~\ref{fig:2}a. The application of an off-resonant laser (detuned by $\Delta$ from NV$_\text{D}$'s $E_x$ transition) for fixed time $T$ during the free precession period $\tau$ projects the NV into the mixed state:
\begin{align}
\rho = \left(1-\Gamma\right)\ketbra{\psi}{\psi}+\sum_k\left(\sum_{ij}\Gamma_{ijk}\ketbra{k}{k}\right),
\end{align}
where $\ket{\psi}=\frac{1}{\sqrt{2}}\left(\ket{0}+e^{-i\theta(t)}\ket{1}\right)$ is the result of the Ramsey experiment and $\sum\Gamma_{ijk}=\Gamma$. In our experiment, the excitation parameters and branching ratios are such that $\Gamma_{0,E_x,0}$ dominates the decay. The summed terms in Equation~2 are stationary states and provide no contrast in the Ramsey experiment, such that the fringe amplitude is directly proportional to $1-\Gamma$. The final spin state (after the second $\pi$/2-pulse of the Ramsey sequence) is measured by state-dependent fluorescence $F$ through 532 nm illumination. $F$ is normalized to account for power fluctuations by repeating the sequence, but replacing the final $\pi/2$-gate with a $3\pi/2$-gate and taking the contrast $C=\frac{(F_{3\pi/2}-F_{\pi/2})}{(F_{3\pi/2}+F_{\pi/2})}$.

Figure~\ref{fig:2}b plots $C$ for varying $\Delta$. For $\Delta=0$, the Ramsey contrast vanishes, as expected for the laser-induced state projection. With increasing detuning, the fringe contrast recovers, approaching a control experiment without readout laser. 

We map the crosstalk as a function of $\Delta$ by fixing the precession time to the fringe maximum at 386~ns and sweeping the resonant laser over a wide range of detunings. These data are converted to a bit error probability in Figure 2c by normalizing the fluorescence from each detuning to that from the reference ``no-laser'' control experiment (Fig. 2b), which gives the crosstalk-free case. The red curve represents our model from Equation~\ref{eq:crosstalk} with only one fit parameter for the optical Rabi frequency, which is difficult to accurately measure experimentally due to spectral diffusion of the ZPL. The optical excitation time $T$ is fixed by our pulse generator, and the decay rate is determined by lifetime characterization~\cite{supplement}. The theory shows good agreement with our data and indicates that a detuning of 16~GHz or greater keeps crosstalk errors below 1\%, a regime accessible by the cluster at site~1 of Figure~\ref{fig:1}c.

We now demonstrate individual control and readout on this cluster. We achieve independent microwave control of the spin states by applying a magnetic field which splits the spin levels depending on the NV center crystal orientation. In this cluster, we find that two of the NV centers (A and B) are oriented along one crystal axis and the third (C) along another, indicated by four dips in the magnetic resonance spectrum~\cite{supplement}. 

\begin{figure}[b]
\centering
\includegraphics{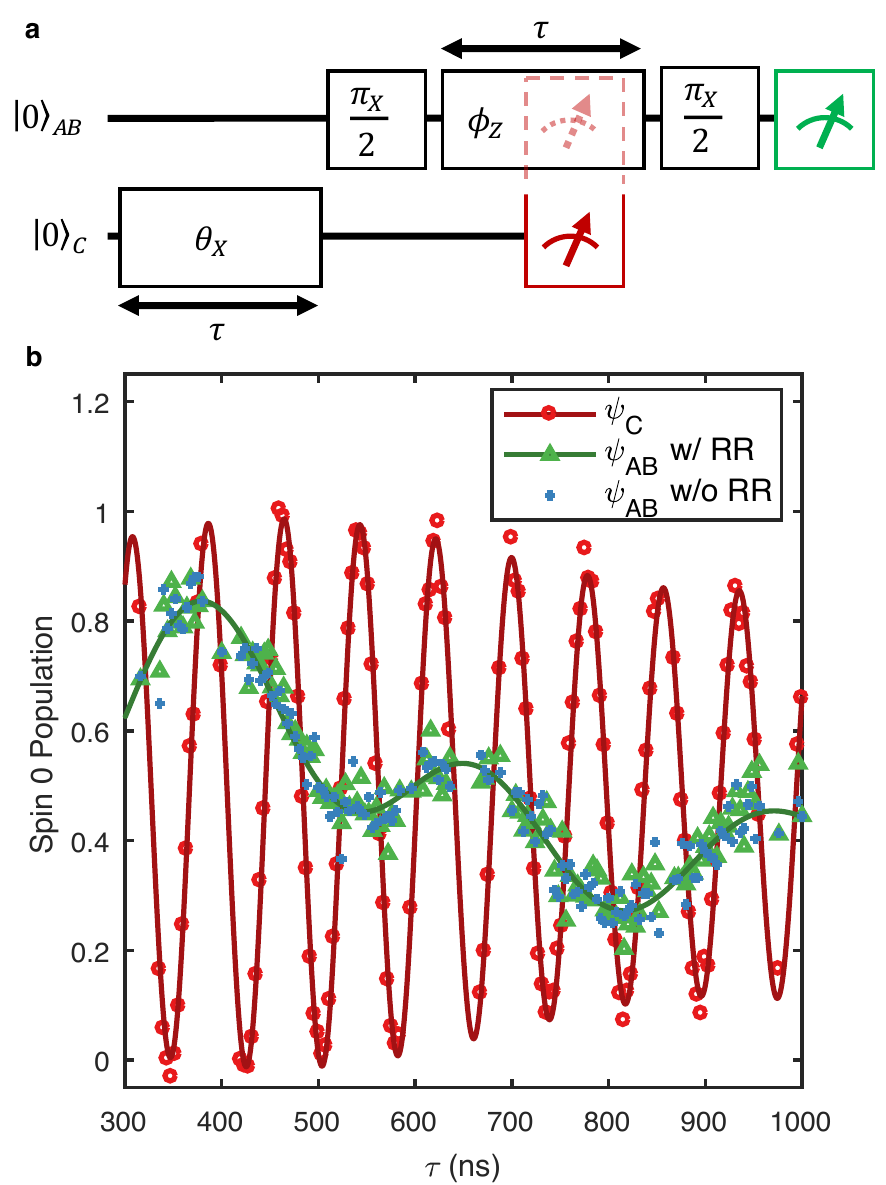}
\caption{\textbf{Simultaneous control and readout of three NVs in a sub-diffraction volume.} \textbf{a}, Gate sequence for demonstrating simultaneous control and readout of multiple NVs. A Rabi sequence is performed on NV C, followed by a subsequent Ramsey sequence on NVs A and B. During the Ramsey sequence, the state of NV C is read out using resonant readout (RR). The RR gate is depicted as partially spilling into $\ket{\psi}_{AB}$ to indicate the possibility of crosstalk. \textbf{b}, Results of the sequence in \textbf{a}, showing Rabi oscillations on NV C (red circles) and Ramsey fringes for NVs A and B (green triangles), alongside the reference Ramsey fringes (blue diamonds) for NVs A and B taken with no Rabi sequence nor RR on NV C. The green fit to the data for $\ket{\psi}_{AB}$ indicates no signal degradation within measurement error bounds~\cite{supplement}.}
\label{fig:3}
\end{figure}

We take advantage of this ground state splitting and apply the same Ramsey sequence from above to perform individual control and readout. Figure~\ref{fig:3}a shows the gate representation of our sequence. After initialization of all three NV centers with a 532~nm repump, the spin of NV C is coherently driven with a resonant microwave pulse for a time $\tau$, inducing Rabi oscillations corresponding to a rotation of angle $\theta$ about the $X$-axis. Next, NVs A and B are rotated into an equal superposition state by a $\pi/2$-pulse, followed by a passive precession by angle $\phi$ about the $Z$-axis for the same time $\tau$. While NVs A and B are in this phase-sensitive superposition state, we perform individual readout on NV C using a resonant optical pulse. After waiting a total precession time $\tau$, a final $\pi/2$-pulse completes the Ramsey sequence on NVs A and B, and we read out these states with 532~nm light~\cite{supplement}. Note that while limitations in the available equipment necessitated the use of a non-resonant green readout on NVs A and B, additional lasers or modulators would allow for individual readout of each NV center in the cluster. Figure~\ref{fig:3}b shows the results of each readout window, where both gates measure the expected Rabi and Ramsey signals. Comparing these Ramsey results to that of a control Ramsey experiment on NVs A and B taken with no additional control or readout sequences on NV C, the fringe amplitudes are equal within our noise bounds (0$\pm$4\% bit error probability). That is, we find no detectable fringe amplitude degradation as a result of the resonant readout pulse, indicating that the states of the off-resonant NV centers are left unperturbed through this readout. This result is consistent with our model, which predicts a bit error probability of ${\sim1}$\%, below the 4\% fit bounds.

\begin{figure*}
\centering
\includegraphics{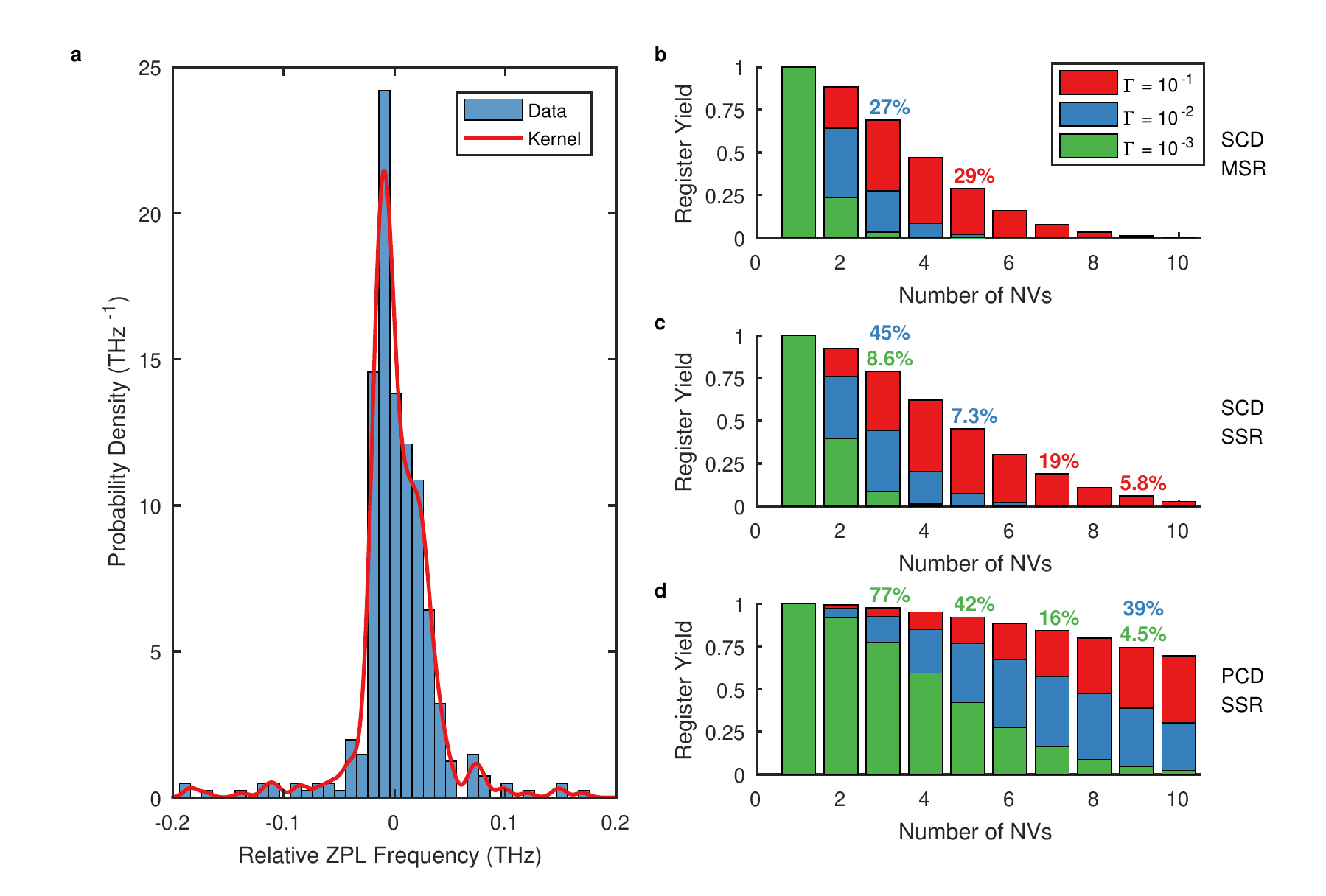}
\caption{\textbf{Architecture scalability.} \textbf{a}, Histogram (blue bars) of 406 zero-phonon line (ZPL) resonance frequencies normalized to be in units of probability density, and corresponding kernel density estimate (red line) of this inhomogeneous distribution. \textbf{b-d}, Simulated probabilities of successfully creating viable registers of varying numbers of NVs under different tolerance thresholds for the probability $\Gamma$ of undesired spontaneous decays from off-resonant NVs. \textbf{b}, Results using MSR parameters shown in this work and the low-strain SCD distribution in \textbf{a}. \textbf{c}, Results using parameters used to demonstrate high-fidelity SSR~\cite{Robledo_2011,Hensen_2015} and the low-strain SCD distribution in \textbf{a}. \textbf{d}, Results using single-shot readout parameters and the high-strain distribution measured in the PCD sample from Fig. 1--3.}
\label{fig:4}
\end{figure*}

We assess the viability of this platform for scalable creation of multi-spin registers by considering the probability of forming systems of multiple distinguishable emitters. Figure~\ref{fig:4}a shows the inhomgeneous distribution of NV center ZPL frequencies acquired on a single-crystal diamond (SCD, see Methods) from a dataset of 406 ZPL transitions from 197 distinct emitter sites. We consider here an SCD to allow comparison to samples most typically used in diamond quantum information experiments~\cite{Waldherr_2014,Hensen_2015,Choi_2017}. From this distribution, we build an empirical kernel estimate (red curve, Figure~\ref{fig:4}a).

Based on Monte Carlo sampling from this measured distribution, we estimate the probability that $N$ emitters in a cluster have a crosstalk probability $\leq\Gamma$ under the parameters used for our multi-shot readout (MSR) in Figure~\ref{fig:2}. These results are given in Figure~\ref{fig:4}b. For example, the probability for an ${N=3}$ NV site to have a crosstalk $\Gamma\leq10^{-2}$ is estimated at 21\%. If each of the NVs were coupled to 3 nuclear spin data qubits, such an $N=3$ system would be sufficient for implementing the [[9,1,3]] Shor-Bacon code~\cite{Bacon_2006}. 

Improved light collection using photonic microstructuring and single-shot readout (SSR)~\cite{Robledo_2011} could markedly improve these yields. To this end, we repeat our simulation under experimental parameters ($\Omega=\gamma$, $T=3.7$~\textmu s) comparable to those used to achieve single shot readout with 97\%~fidelity in a solid immersion lens~\cite{Hensen_2015}. The results in Figure~\ref{fig:4}c indicate a twofold increase in the $N=3$ yield discussed above. Keeping these readout parameters and additionally assuming the measured ZPL distribution for the PCD (Figure~\ref{fig:1}d) produces the yield histogram in Figure~\ref{fig:4}d, which shows that registers with $N=9$ NVs could be produced with 39\% yield with $\Gamma\leq10^{-2}$.

In conclusion, we demonstrated readout of individual solid-state qubits within a diffraction-limited cluster with crosstalk on non-addressed qubits below $4\times10^{-2}$. This capability was enabled by strain-splitting of the NV centers' ZPL transitions in a PCD. While this work uses native strain, the strain field may also be engineered~\cite{Meesala_2018} to provide greater control and increase inhomogeneous distributions. If an application necessitates a low-strain environment, these resonances can also be shifted by applying a DC electric field, allowing defects of different orientations --- or the same orientation under a strong field gradient --- to be uniquely addressed. The approach presented here is also applicable to other atom-like emitters, such as quantum dots~\cite{Press_2008}, rare-earth ions~\cite{Dibos_2018}, and other solid-state color centers. When combined with existing techniques for producing sub-diffraction clusters via aperture implantation~\cite{Jakobi_2016}, entangling defect centers with ancilla nuclear spins~\cite{Abobeih_2018}, and single-shot readout~\cite{Robledo_2011}, this provides a path towards creating large ensembles of individually-addressable qubits, with a number of applications. For example, error-corrected registers using the 7-qubit~\cite{Steane_1996} or 9-qubit~\cite{Bacon_2006} codes could be constructed with clusters of multiple coupled NV-dark spin systems, with full connectivity given by spin-spin coupling between adjacent NV centers. This would allow extension of architectures comprising one optically active and multiple dark spins~\cite{Abobeih_2018} by reducing the problem of spectral crowding, as well as increasing the effective gate rate by parallelization. NV clusters with individual readout could also enable entanglement-assisted and spatially-resolved nanoscopic quantum sensing~\cite{Degen_2017}. Finally, such clusters present an appealing architecture for modular quantum computing schemes~\cite{Nickerson_2014,Pant_2017} and spectrally-multiplexed quantum repeaters~\cite{Sinclair_2014}.

\section*{Methods}
\subsection*{Sample preparation}
Super-resolution localization and individual control experiments were performed on a Type IIa PCD by chemical vapor deposition (Element 6), with a native nitrogen concentration of $<$50~ppb. To increase the prevalence of sub-diffraction clusters, the sample was implanted with nitrogen at 85~keV with a density of 10$^{10}$~cm$^{-2}$, and subsequently annealed at 1200\textdegree~C for 8 hours. The conversion yield of this process is on the order of 1\%, resulting in a final areal NV density of roughly 1~\textmu m$^{-2}$. For delivering microwaves, striplines of 100~nm thick gold with a 5 nm titanium adhesion layer were fabricated via lift-off on the diamond surface. The single crystal Type IIa diamond was also produced by chemical vapor deposition (Element 6), with a native nitrogen concentration of $<$5 ppb. NV centers in this sample were created by implanting nitrogen (energy 185~eV, dosage 10$^9$~cm$^{-2}$) and subsequently annealing at 1200\textdegree~C for 8 hours.

\subsection*{Experimental set-up}
Experiments were performed using a home-built scanning confocal microscope. The samples were cooled to 4 K using a closed-cycle helium cryostat (Montana Instruments) and were imaged through a 0.9~NA vacuum objective. 532 nm light was generated by a Coherent Verdi G5 laser, and resonant red light tunable around 637 nm was generated by a New Focus Velocity tunable diode laser. Microwave signals were generated by Rohde \& Schwarz SMIQ06B and SMV03 signal generators and sent through a high-power amplifier (Mini-Circuits ZHL-16W-43+) before delivery to the sample.

\subsection*{Data availability}
The data that support the plots within this paper and other findings of this study are available from the corresponding author upon reasonable request.

\bibliography{references}

\section*{Acknowledgements}
E.B. was supported by a NASA Space Technology Research Fellowship and the NSF Center for Ultracold Atoms (CUA). M.W. was supported by the STC Center for Integrated Quantum Materials (CIQM), NSF Grant No. DMR-1231319, the Army Research Laboratory Center for Distributed Quantum Information (CDQI), and Master Dynamic Limited. S.L.M. was supported by the NSF EFRI-ACQUIRE program Scalable Quantum Communications with Error-Corrected Semiconductor Qubits and the AFOSR Quantum Memories MURI. M.E.T. was supported by an appointment to the Intelligence Community Postdoctoral Research Fellowship Program at MIT, administered by Oak Ridge Institute for Science and Education through an interagency agreement between the U.S. Department of Energy and the Office of the Director of National Intelligence. T.S. was supported by the European Union’s Horizon 2020 research and innovation programme under the Marie Skłodowska-Curie grant agreement no. 753067 (OPHOCS). This work was supported in part by the AFOSR MURI for Optimal Measurements for Scalable Quantum Technologies (FA9550-14-1-0052) and by the AFOSR program FA9550-16-1-0391, supervised by Gernot Pomrenke. The authors would like to thank C. Foy, H. Choi, and C. Peng for helpful discussions.

\section*{Author Contributions}
E.B. and M.W. performed the experiments. E.B. and T.S. constructed the optical setup. S.L.M., M.E.T., and M.W. prepared the sample. E.B., M.W., M.E.T., T.S., and D.R.E. conceived the experiments. E.B., M.W., S.L.M., and D.R.E. prepared the manuscript. All authors reviewed the manuscript.

\section*{Competing Financial Interests}
The authors declare no competing financial interests.

\end{document}